\documentclass[journal=acsphotonics,manuscript=article]{achemso}

\usepackage[version=3]{mhchem} 

\author{Iago Diez}
\email{ir274@exeter.ac.uk}
\affiliation[University of Exeter]
{College of Engineering, Mathematics and Physical Sciences, University of Exeter, EX4 4QF, Exeter, United Kingdom}
\author{Andrey Krysa}
\affiliation[University of Sheffield]
{EPSRC National Epitaxy Facility, University of Sheffield, S1 3JD, Sheffield, United Kingdom}
\author{Isaac J. Luxmoore}
\email{i.j.luxmoore@exeter.ac.uk}
\affiliation[University of Exeter]
{College of Engineering, Mathematics and Physical Sciences, University of Exeter, EX4 4QF, Exeter, United Kingdom}

\title[]
  {Inverse Design of Whispering-gallery Nanolasers with Tailored Beam Shape and Polarization}

\abbreviations{WGM, QW, TO, FoM, AP, RP, LP}

\begin{document}
\begin{tocentry}
\begin{figure}[H]
\centering
\includegraphics[width=0.81\textwidth]{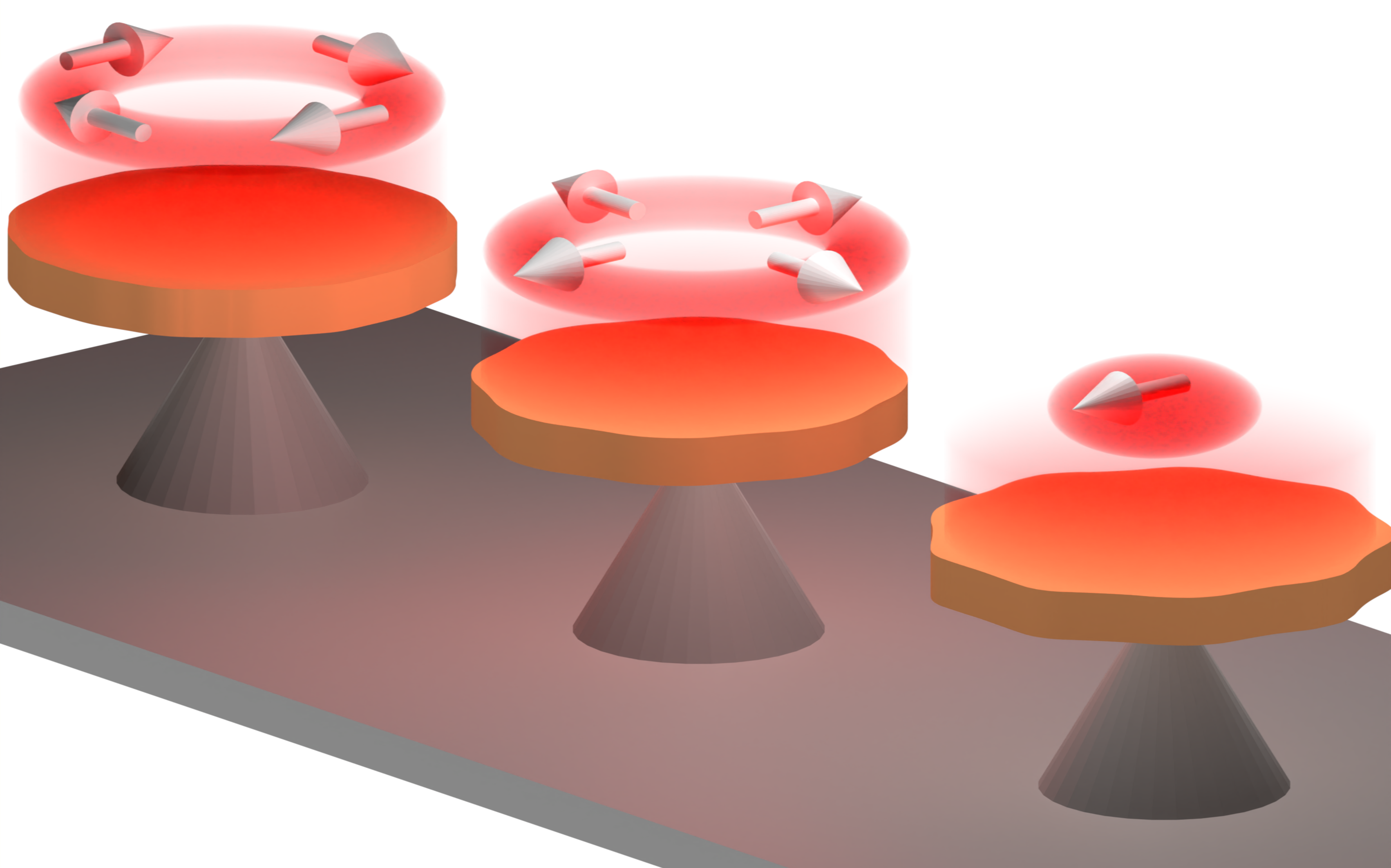}
\end{figure}
\end{tocentry}


\begin{abstract}
Control over the shape and polarization of the beam emitted by a laser source is important in applications such as optical communications, optical manipulation and high-resolution optical imaging.
In this paper, we present the inverse design of monolithic whispering-gallery nanolasers which emit along their axial direction with a tailored laser beam shape and polarization.
We design and experimentally verify three types of sub-micron cavities, each one emitting into a different laser radiation mode: an azimuthally polarized doughnut beam, a radially polarized doughnut beam and a linearly polarized gaussian-like beam. The measured output laser beams yield a field overlap with respect to the target mode of 92\%, 96\% and 85\% for the azimuthal, radial and linearly polarized cases, respectively, thereby demonstrating the generality of the method in the design of ultra-compact lasers with tailored beams.
\end{abstract}

\section{Introduction}
Spatial control over the shape and polarization of the beam emitted by a laser source is becoming increasingly relevant for applications such as polarization multiplexing in optical communications \cite{Milione2015, Ivanovich2020}, stiffer optical trapping \cite{Michihata2009} and high-resolution optical imaging \cite{Kozawa2018}. This tailoring of the beam is conventionally done with optics external to the source, but recently there has been interest in structuring light at the source\cite{Forbes2019}. This is specially relevant to meet the growing demand for higher density photonic integration and for further miniaturization of laser sources\cite{Zhou2015b,Wu2019}. However, on-chip generation of tailored laser beams has not yet been realized for cavities under the sub-micron footprint.

One of the most compact types of integrated cavities are the optical whispering-gallery mode microdisc laser, which exhibit large quality factors and low mode volumes, thereby enhancing the light-matter interaction resulting in low lasing threshold. These properties make them excellent systems for applications such as biochemical sensing \cite{Toropov2021,Fang2004}, cell barcoding and tracking\cite{Fikouras2018,Tang2021} and optical communications \cite{Zhukov2021}. However, due to their circular geometry, the laser light is radiated in-plane and isotropically \cite{Tang2021,Yang2015}. This limits their range of use due to the poor collection efficiency in the axial direction and to date there has been little effort to engineer the emission beam shape and/or polarization. 

Previous work to achieve vertical emission from WGM lasers has focused on the addition of an angular grating that scatters the light out of the plane. The gratings have been fabricated by metal deposition on top of the microdisc cavity \cite{Mahler2009}, by etching at the inner wall of a microring cavity \cite{Zhang2020TunableMicrolaser,Shao2018} or at the outer wall of a  microdisc cavity \cite{Al-Attili2018}. In all these examples, the cavities present a lateral footprint larger than 2 \(\mu\)m in diameter and polarization control over the radiated beam was only demonstrated for a microring cavity of 120 \(\mu\)m diameter \cite{Shao2018}.

In this work, we present an alternative approach, whereby an inverse design algorithm is applied to optimize a sub-micron scale WGM-cavity design for axial emission, with on-demand laser beam shape and polarization. Inverse design is a powerful tool for discovering novel geometries and for optimizing performance, which is not reliant on the intuition of the user. The inverse design process follows an iterative optimization strategy that allows exploration of the whole design parameter space and has been shown to be effective in a variety of applications, including: maximization of band gaps in photonic crystals \cite{Kao2005MaximizingMethods}; wide-angle diffractive optical elements \cite{Kim2020InverseElements}; photon extractors for nitrogen-vacancy centers \cite{Chakravarthi2020,Wambold2020Adjoint-optimizedDiamond}; power splitters in integrated photonics \cite{Piggott2017Fabrication-constrainedDesign};  coupling of an optical antenna-LED to a single-mode waveguide \cite{Andrade2019InverseWaveguide} and on-chip resonators\cite{Ahn2022}.

Here, we implement a topology optimization (TO) algorithm \cite{Molesky2018,Lalau-Keraly2013AdjointDesign} for tailoring the laser beam shape and polarization. To show the generality of the method we apply it to the design of three WGM-cavities, each one emitting into a different radiation mode: two spatially varying polarization doughnut beams with azimuthal (AP) and radial (RP) polarization; and a spatially homogeneous linearly polarized (LP) gaussian-like beam. The three different designs are achieved for the same WGM and for a wavelength in the range 650 -- 700 nm (chosen to coincide with the QW gain spectral range). The nanocavity designs are fabricated from a GaInP/AlGaInP double quantum well (QW) wafer and their lasing and far-field verified by Fourier microscopy, k-space polarimetry and photoluminescence spectroscopy.

\section{Inverse Design Problem}
\begin{figure}[hbtp]
\centering
\includegraphics[width=1\textwidth]{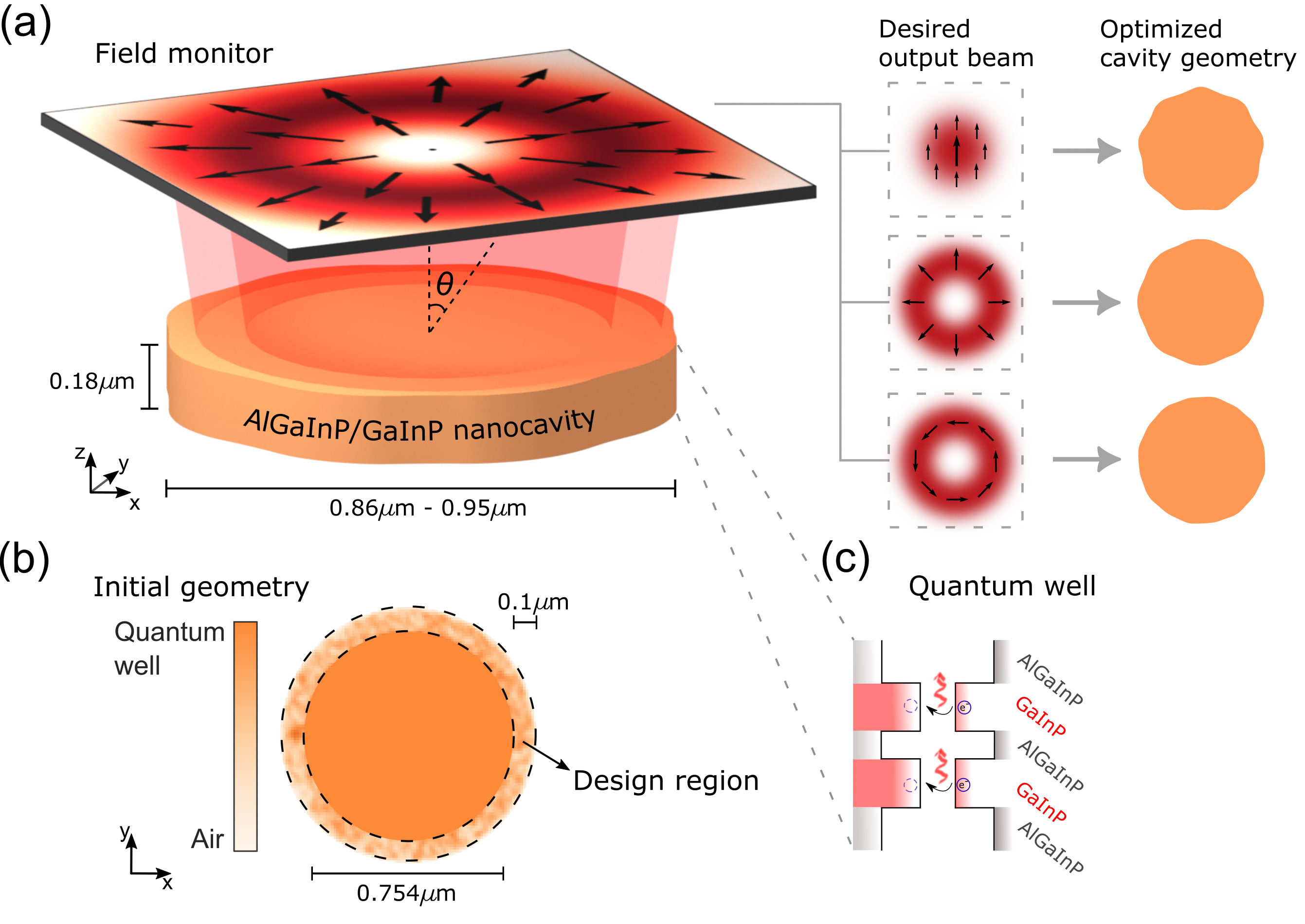}
\caption{(a) Schematic of the inverse-designed nanocavity showing the monitor that records the electric fields in free space. The inverse design algorithm optimizes the nanocavity geometry in order to realize the on-demand beam shape and polarization. (b) The design region is an annulus surrounding an inner disc made of quantum well material. The design region is initialized with a randomized distribution of dielectric permittivity values. (c) The band structure schematic of the double quantum well GaInP/AlGaInP shows the composition of the well and barrier layers.}
\label{fig:schematicnanolaser}
\end{figure}

 The inverse design of the nanocavity geometry is performed using an adjoint-based topology optimization (TO) method, illustrated schematically in Figure~\ref{fig:schematicnanolaser}. TO is a computational technique for inverse design that can handle extensive design spaces considering the dielectric permittivity at every spatial point as a design variable. The algorithm maximizes the overlap between the mode radiated by the cavity, \(\mathbf{E}\), and the desired radiation mode, \(\mathbf{E}_m\) in free space: \(\sim \left| \mathbf{E}_m^* \cdot  \mathbf{E} \right|\); this overlap is the Figure of Merit (FoM) or objective function of our TO problem. This approach gives control over the final output beam shape and polarization by allowing the user to select an appropriate target mode \(\mathbf{E}_m\). The adjoint method allows the efficient computation of the optimization gradient, which indicates how to evolve the cavity geometry with only two electromagnetic simulations, known as the \textit{forward} and \textit{adjoint} simulations, regardless of the number of degrees of freedom \cite{Molesky2018,Lalau-Keraly2013AdjointDesign,Jensen2011,Miller2013,Giles2000AnDesign}. The electromagnetic simulations are computed using a finite-difference time-domain Maxwell's equation solver (Lumerical). 

The iterative optimization procedure is initiated with the geometry shown in Figure~\ref{fig:schematicnanolaser}(b), consisting of an inner disk of diameter 754 nm, surrounded by an external 100 nm wide annulus, both with a fixed thickness of 180 nm. The dielectric permittivity of the inner disc was fixed at $\varepsilon_{QW}=11.56$, by the QW wafer, and the annulus had an initial random distribution of dielectric permittivity ranging from $\varepsilon_{air} =1$ to $\varepsilon_{QW}$, with air being the background medium. The optimization is performed only for the annulus in order to keep the cavity structure as a continuous object, i.e. without holes. This initial disc cavity supports a WGM of order 8 in the azimuthal direction and fundamental order along the radial and axial direction, for the target wavelength of $\sim670~\mathrm{nm}$.

The simulation space was parametrized with the density parameter $\rho$, which is a linear transformation on the dielectric permittivity $\varepsilon$ $\epsilon$ $\left[\varepsilon_{air},\varepsilon_{QW}\right]$ so that the density values range from 0 to 1, i.e. $\rho$  $\epsilon$  $[0, 1]$ \cite{Jensen2005TopologyWaveguide}.
\begin{equation}
    \rho \left( \varepsilon \left( \mathbf{x}\right) \right) = \frac{\varepsilon\left( \mathbf{x}\right)-\varepsilon_{air}}{ \varepsilon_{QW}-\varepsilon_{air}}
\end{equation}
The spatial resolution of the parametrization was 10 nm in the three directions $X$, $Y$, and $Z$.

The steps of the iterative TO process are as follows:
\begin{enumerate}
\item Density filtering \cite{Sigmund2007,Jensen2011} is applied to the density distribution \(\rho\) to avoid the presence of sharp features that would be beyond practical fabrication capability: \(\rho \rightarrow \tilde{\rho} \). More information about the spatial filtering can be found in the Supporting Information (S.I.).
      
\item The forward simulation is run. The source used for exciting the WGM inside the cavity was an electric dipole polarized along the radial direction of the disc and placed close to the disc contour, where an antinode of the WGM would be expected. The electric field $\mathbf{E}_{fwd}$ throughout the design region, and the electric field $\mathbf{E}$ throughout the FoM plane (field monitor on Figure~\ref{fig:schematicnanolaser}(a)) are recorded. The FoM plane is placed at a distance of \(\sim0.720\) \(\mu\)m above the top surface of the cavity. We compute the FoM value \(\mathcal{F}\) with:

\begin{equation}
    \label{eq:FOM_em}
    \mathcal{F} =  \frac{\left| \int \mathbf{E}_m^* \cdot  \mathbf{E} dxdy \right|}{\sqrt{\int \left|\mathbf{E}_m\right|^2 dxdy} \sqrt{\int \left|\mathbf{E
    }\right|^2 dxdy}}
\end{equation}
    
\item The adjoint simulation is run. The adjoint source is a mode launched from free space above the cavity with the conjugate of the desired spatial distribution of intensity and polarization: $\mathbf{E}_m^*$. At the end of the simulation, the electric field $\mathbf{E}_{adj}$ throughout the design region is recorded.

The resonant wavelength of the particular WGM of interest shifts at each iteration due to the variation in the dielectric distribution within the design region, and therefore must be tracked. The wavelength at which $\mathbf{E}_{fwd}$ and $\mathbf{E}_{adj}$ are recorded is therefore updated at each iteration according to the maximum of the Purcell enhancement.

\item The gradient of the FoM with respect to the density distribution \(\tilde{\rho}(\mathbf{x})\) is calculated \cite{Lalau-Keraly2013AdjointDesign} as: \(G = Re \{ \mathbf{E}_{fwd} \cdot \mathbf{E}_{adj}\}\). This gradient indicates whether the dielectric permittivity of each point within the design region should be increased (\(G>0\)) or decreased (\(G<0\)),  in order to maximize \(\mathcal{F}\).
    
\item Finally, the density distribution is updated: $\hat{\rho} = \tilde{\rho} + \alpha G/max|G|$, where $\alpha$ is a hyperparameter that controls the evolution rate, but taking into consideration that the new density value is bounded: $\hat{\rho}$ $\epsilon$ $[0,1]$. In our simulations we chose $\alpha = 0.05$ to approach to infinitesimal changes in the dielectric permittivity while keeping a large enough value to maintain a reasonable run-time for the optimization.

\end{enumerate}
   
This process is repeated for 100 iterations to check the stability of the optimized design, but can be stopped as soon as \(\mathcal{F}\) converges to a value. The final designs are obtained by binarizing the density distribution \(\hat{\rho}\) from the last iteration with the threshold being 0.5: \[\rho_{bin} = 1 \quad \text{where} \quad \hat{\rho}>=0.5 \quad \quad, \quad\quad  \rho_{bin} = 0 \quad \text{where} \quad \hat{\rho}<0.5\]

\section{Results and Discussion}
The TO procedure is used to design cavities that radiate into the three different target output beams: azimuthally polarized doughnut (AP), radially polarized doughnut (RP) and linearly polarized gaussian-like (LP). The resulting optimized designs are shown in Figure~\ref{fig:schematicnanolaser}(a).

\begin{figure}[hbtp]
\centering
\includegraphics[width=1\textwidth]{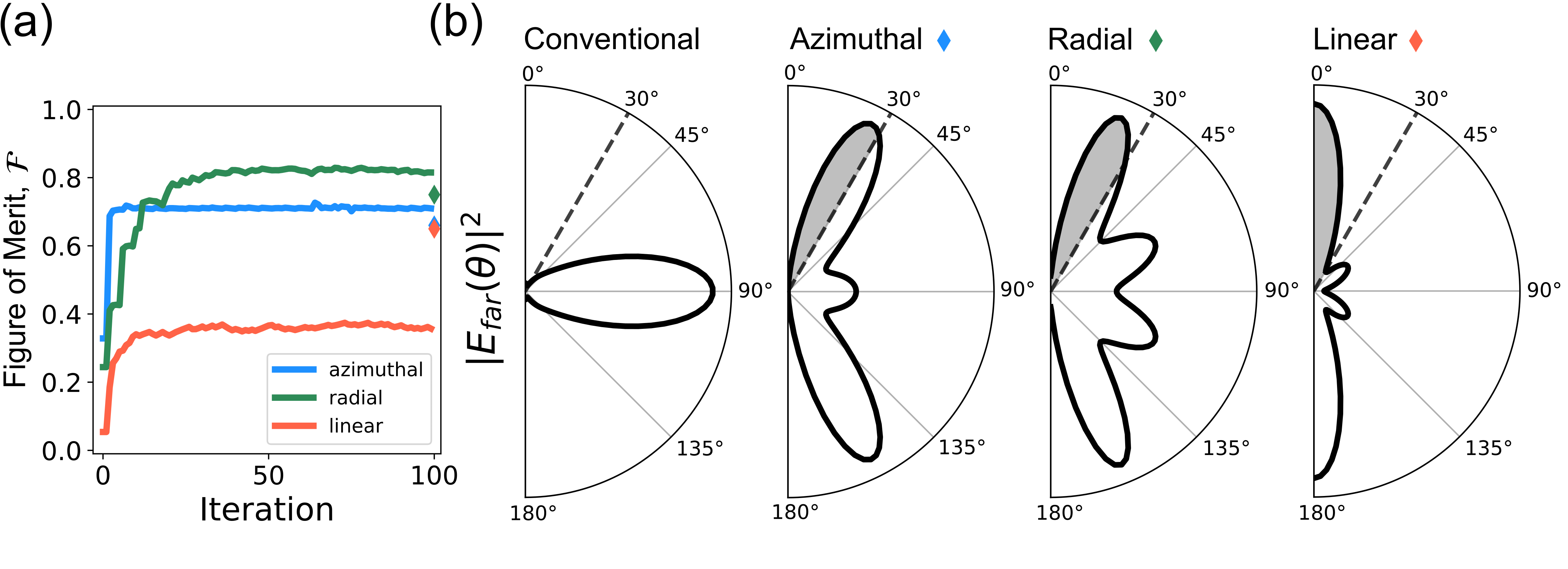}
\caption{(a) The evolution of the figure of merit, \(\mathcal{F}\), during the topology optimization process. The last value, represented by a rhombus of the same color as the solid lines, corresponds to the binarized structure. (b) The far-field intensity \(\left| E_{\text{far}}\right|^2\), integrated along the azimuthal angle, is plotted against the polar angle \(\theta\), for the conventional circular cavity and the inverse-designed cavities optimized for the radiation modes with azimuthal (AP), radial (RP) and linear (LP) polarization. The intensity that would be collected by an objective lens of numerical aperture 0.5 (maximum acceptance angle 30\(^{\circ}\)) is represented by the gray shaded area. }
\label{fig:FoM_Epolar}
\end{figure}

The convergence to the optimized design can be seen in the evolution of \(\mathcal{F}\) in Figure~\ref{fig:FoM_Epolar}(a). For the three designs, \(\mathcal{F}\) increases gradually up to a saturation value which is stable within \(\leq0.005\). This saturation indicates the TO algorithm has found a local optimum solution for the optimization problem, given the initial density distribution and constraints such as the design region size and the spatial filtering. \(\mathcal{F}\) for the binarized geometries is plotted with a rhombus of the same color as the respective solid lines. There is a drop in value for AP and RP binarized cavities, whereas there is an increase in value for the LP cavity. Figure~\ref{fig:FoM_Epolar}(a) shows the evolution of \(\mathcal{F}\) evaluated by using the \textit{near-field} recorded at the field monitor. Even though the algorithm is maximizing the near-field overlap, the \textit{far-field} overlap increases accordingly. The evolution of \(\mathcal{F}\) computed with a far-field projection is shown in the S.I. 

As a consequence of optimizing for a target radiation mode above the cavity, the power radiated out of the plane is increased. This effect can be seen in the radiation diagrams of the optimized cavities plotted in Figure~\ref{fig:FoM_Epolar}(b). These plots show the far-field intensity radiated at each polar angle \(\theta\), upon integration of all azimuthal angles. As discussed in the introduction, the conventional circular cavity radiates most of its power in the plane of the cavity, with a maximum intensity at \(\theta=90^{\circ}\). In comparison, the optimized designs have maximum intensity at \(\theta=\) 26.3\(^{\circ}\) (AP), 22.3\(^{\circ}\) (RP) and 0\(^{\circ}\) (LP). Furthermore, the radiated intensity transmitted to a hypothetical objective lens, of numerical aperture 0.5 (maximum acceptance angle 30\(^{\circ}\)), placed above the cavity is also increased. As summarized in Table~\ref{tbl:results_simulations}, the fraction of intensity that is collected by such objective, i.e. the transmitted power \(T\), is 21.8\% (AP), 20.8\% (RP) and 33.7\% (LP). Compared to the 2.5\% that would be collected from a conventional circular cavity, this corresponds to an enhancement in power collection of \(\times\)8.7, \(\times\)8.3 and \(\times\)13.5, respectively.

\begin{table}
  \caption{Performance results from the simulation of the optimized nanocavities shown in Figure~\ref{fig:FoM_Epolar}(a). For each radiation mode the table shows the fraction of power transmitted within this angular aperture and the expected enhancement in power collection when compared to the radiation from a conventional circular cavity.}
  \label{tbl:results_simulations}
  \begin{tabular}{lll}
    \hline
    Cavity type & \(T\) (\(\theta\leq30^{\circ}\)) &  Power collection enhancement \\
    \hline
    Conventional & 2.5 \% &  \(\times\)1 \\
    Azimuthal   &     22 \% & \(\times\) 8.7 \\
    Radial      &     21 \%  &  \(\times\) 8.3 \\
    Linear      &     34 \% &  \(\times\) 13.5 \\
    \hline
  \end{tabular}

\end{table}

\subsection{Experimental Verification}
\begin{figure}[hbtp]
\centering
\includegraphics[width=1\textwidth]{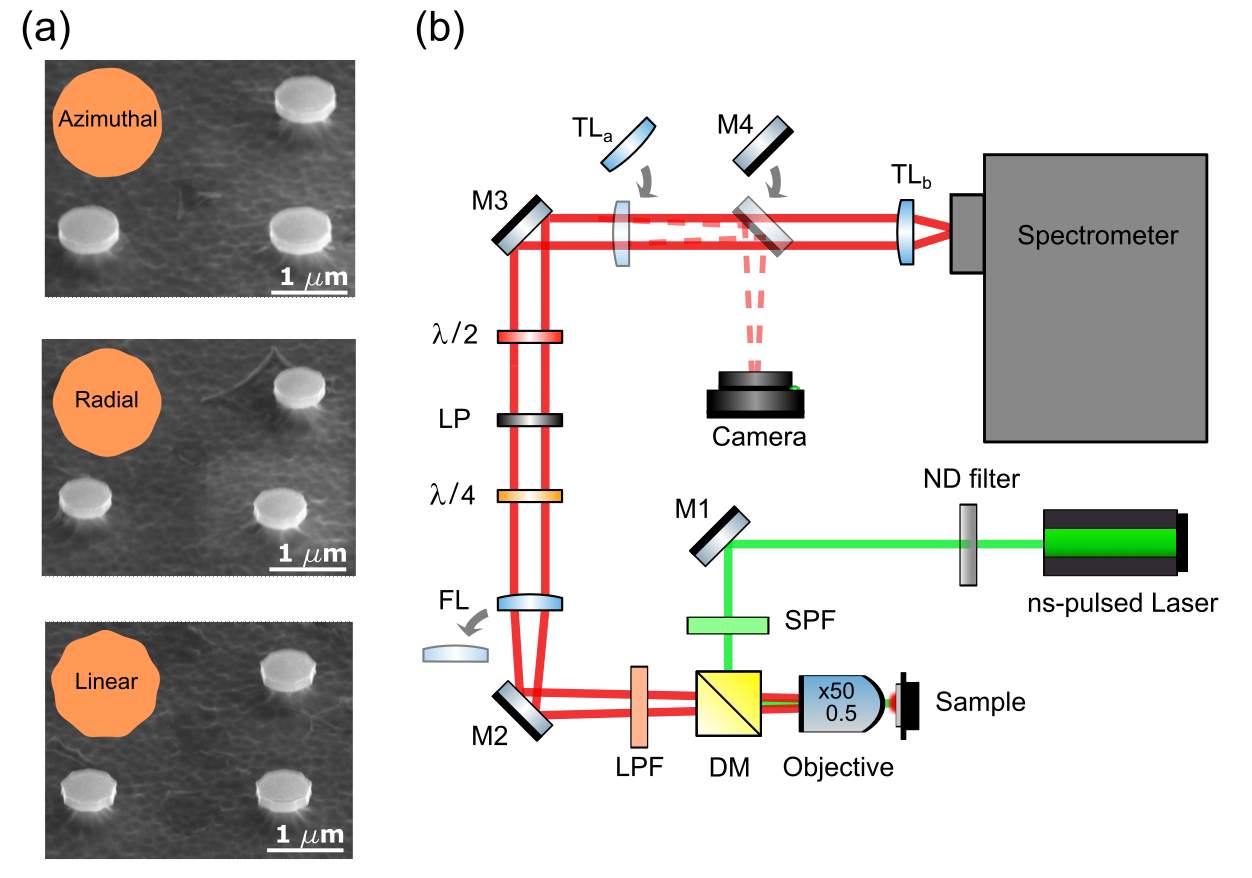}
\caption{(a) Optimized designs and SEM images of nanolasers with azimuthal, radial and linear radiation modes. (b) Diagram of the optical setup to measure the far field intensity and polarisation emitted by the nanolasers and their spectra. The configuration shown in the schematic was used for recording the far-field images with Fourier microscopy. For obtaining real space images of the sample, the collected light was projected onto the camera by removing FL and adding TL\(_a\) and M4 into the beam path.To acquire spectra FL was removed from the beam-path. ND: neutral density, SPF: shortpass filter, DM: dichroic mirror, M\(\#\): dielectric mirror, LPF: longpass filter, FL: Fourier lens, \(\lambda/4\): quarter-wave plate, LP: linear polarizer, \(\lambda/2\): half-wave plate, TL: tube lens.}
\label{fig:opticalSetup}
\end{figure}

To validate the inverse design approach, we fabricate nanolasers from a GaInP/AlGaInP wafer (simple schematic of the QW in Figure~\ref{fig:schematicnanolaser}(c); full layer structure in S.I.) in the shape of the three different optimized cavity designs (Figure~\ref{fig:opticalSetup}(a)) via electron-beam lithography followed by dry and wet etching processes. The nanolasers are characterized using the experimental setup shown schematically in Figure ~\ref{fig:opticalSetup}(b), and assessed against the following criteria: the overlap between the experimental and desired far-field; the maximum output power; the lasing threshold; and the quality factor. The experimental far-field intensity distributions are obtained by Fourier microscopy and their polarization distribution measured with k-space polarimetry \cite{Osorio2015K-spaceAntennas}. The maximum output power, the lasing threshold and the quality factor are obtained from photoluminescence spectra at different excitation powers. Further detail on the fabrication procedure and optical characterization can be found in the Methods section.

To characterize the nanolasers far-field and polarization, we measure the spatial distribution of the Stokes parameters: \(S_0\) is the total intensity; \(S_1\) represents the balance between linearly polarized light intensity at 0\(^{\circ}\) and 180\(^{\circ}\); \(S_2\) refers to the balance between linearly polarized light intensity at 45\(^{\circ}\) and 135\(^{\circ}\); and \(\psi\) represents the polarization ellipse orientation \cite{Osorio2015K-spaceAntennas}. With this information it is possible to recreate the polarization (orientation) of the electric field in the far-field \( [\text{P}_x,\text{P}_y] \), except for its phase. Further details can be found in the S.I.. In the experiment, the Stokes parameters are presented in Fourier space (\(k_x,k_y\)), where the direction of emission is given by the polar angle obtained as \(\theta= \arcsin\left(\sqrt{k_x^2+k_y^2}/k_0\right)\). The Stokes parameters of the target mode are calculated by a far-field projection of the target mode over a hemispherical surface. The target far-field is presented as a top view of this surface with the direction cosines as coordinates (\(u_x,u_y\)), where the polar angle is obtained as \(\theta= \arcsin\left(\sqrt{u_x^2+u_y^2}\right)\))

\begin{figure}[hbtp]
\centering
\includegraphics[width=0.9\textwidth]{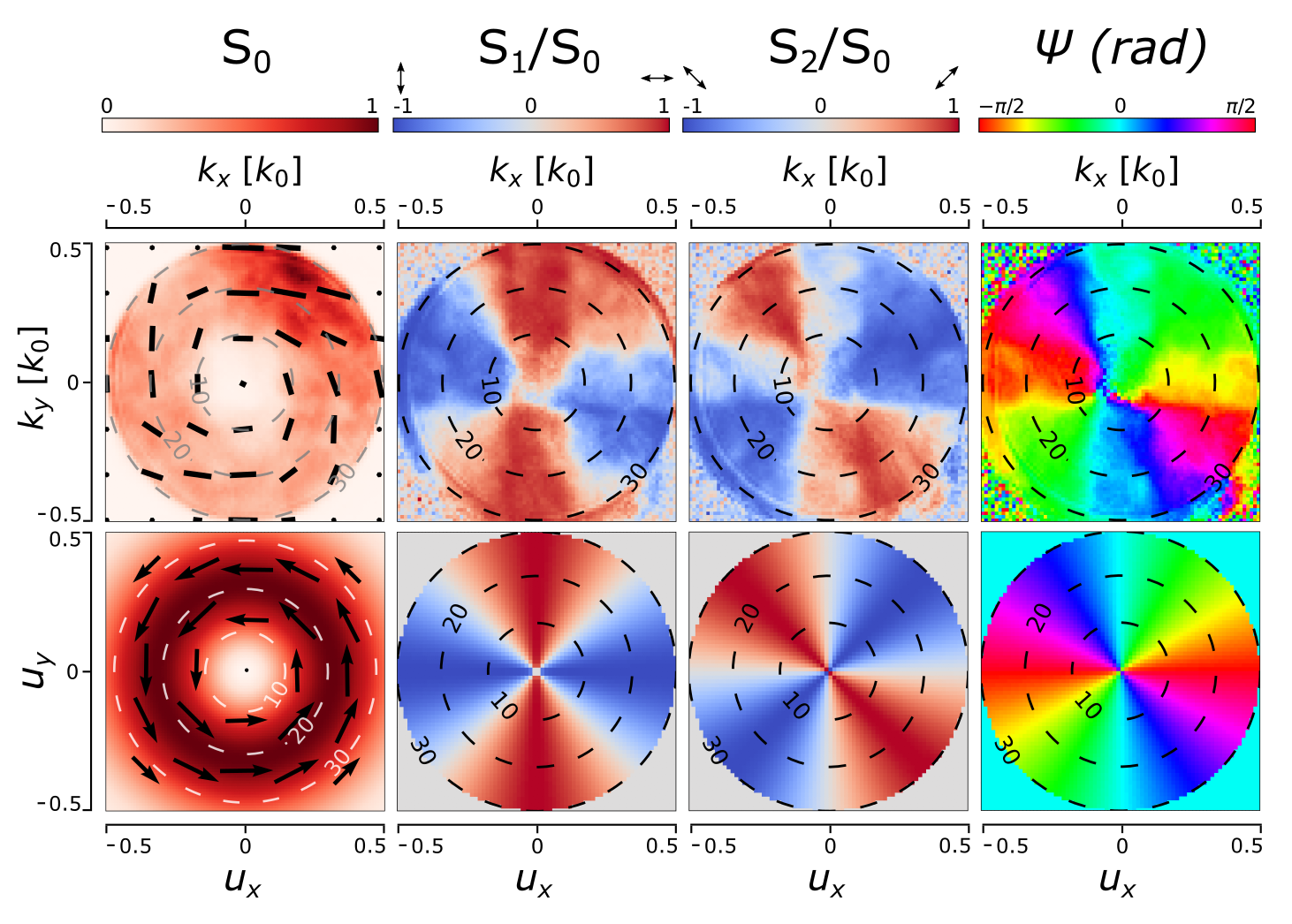}
\caption{Stokes parameters of one of the nanolasers emitting an azimuthally polarized beam. The first row of graphs describe the Stokes parameters of the experimental far-field intensity and polarization in Fourier space; the coordinates \(k_x,k_y\) refer to the wavevectors along the X and Y directions, respectively. The second row of graphs describe the Stokes parameters of the target far-field intensity over a hemisphere; the coordinates \(u_x,u_y\) refer to the direction cosines for the X and Y directions, respectively. The first three Stokes parameters \(S_0\), \(S_1\) and  \(S_2\) and the polarization ellipse angle \(\Psi\) are plotted in colormaps.
The gray, black or white dashed concentric circumferences represent the polar angle and the black arrow-map on the \(S_0\) graph represents the polarization of the electric field.}
\label{fig:detailedStokes}
\end{figure}

Figure~\ref{fig:detailedStokes} presents the experimentally measured spatial distribution of the Stokes parameters for an AP-nanolaser, in comparison to the that of the simulated target AP beam. The experimental \(S_0\) shows the expected doughnut shape with an almost uniform intensity distribution and a null-field at \(\theta<10^{\circ} \;(k_x,k_y\sim0)\) as a consequence of its polarization singularity, thus corroborating the vortex nature of the beam\cite{Zhou2021}. The vectorial nature of the beam is confirmed by the inhomogeneous polarization distribution shown over \(S_0\). \(S_1\), \(S_2\) and \(\Psi\) are in good agreement with the expected distribution but slightly rotated. This rotation is translated into the beam not being purely azimuthally polarized but actually having a small radial component \cite{Moh2007GeneratingBeams}.

\begin{figure}[hbtp]
\centering
\includegraphics[width=1\textwidth]{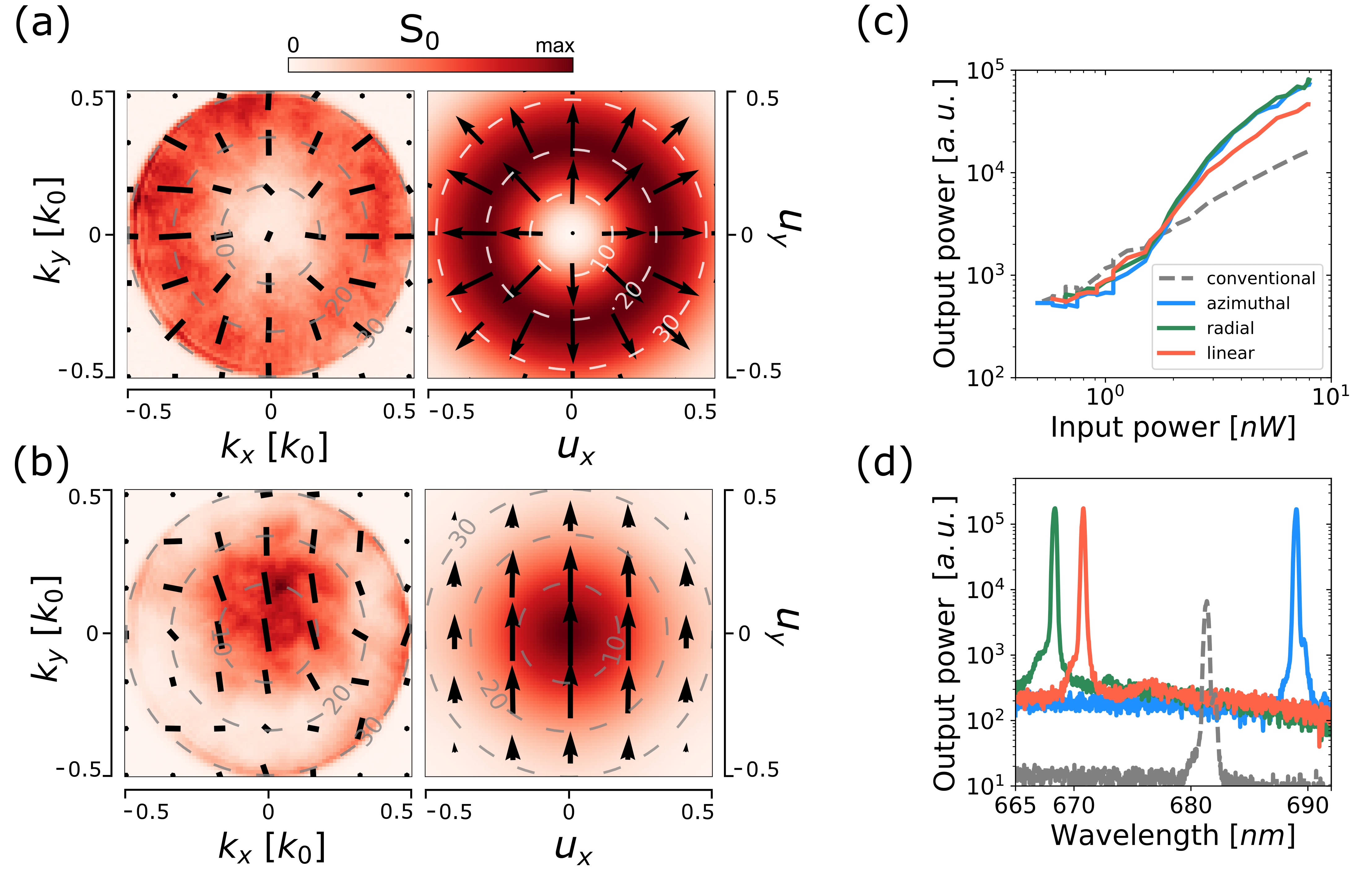}
\caption{Far-field intensity (\(S_0\)) and polarization (black arrow-map) distribution for a radial cavity in (a) and for a linear cavity in (b). The experimental far-field in Fourier space (left column) is placed next to the desired far-field projected on a hemisphere (right column) for visual comparison.
The coordinates \(k_x,k_y\) refer to the wavevectors along the X and Y directions, respectively. 
The coordinates \(u_x,u_y\) refer to the direction cosines for the X and Y directions, respectively.
The gray, black or white dashed concentric circumferences represents the polar angle \(\theta\) in the graphs.
(c) Input-output light power (L--L) curve of a nanolaser that is representative of the ensemble for each type of cavity. The input power values are the averaged power of the pulsed excitation (5 ns, 100 Hz) over a cycle. (d) Photoluminescence spectrum of the nanolasers shown in (c), with the same legend.}
\label{fig:RP_LP_LL_Q}
\end{figure}

A similar comparison is made for RP- and LP-nanolasers, with the experimental and simulated \(S_0\) presented in Figure~\ref{fig:RP_LP_LL_Q} and a full analysis in the S.I. For the RP-nanolaser, \(S_0\) shows the expected doughnut shape with a central null-field and radial polarization throughout (Figure~\ref{fig:RP_LP_LL_Q}(a)). For the LP beam, a gaussian-like intensity distribution is obtained; as shown in Figure~\ref{fig:RP_LP_LL_Q}(b). The center of the beam is more intense than the periphery and the majority of the intensity is collected within a cone of \(\theta \leq20^{\circ}\). The beam is off the optical axis (Z-axis) by \(\sim5^{\circ}\)--\(10^{\circ}\). Also the null-field or polarization singularity of the RP beam seems to be off axis. This might be due to a small misalignment of the sample stage relative to the optical axis of the objective lens.

A quantitative assessment of the experimentally obtained far-field modes is calculated by overlapping the experimental far-field \( [\text{P}_x,\text{P}_y]\) with that of the target mode \( [\text{E}_{m,x},\text{E}_{m,y}] \) for each nanolaser. The three different far-field modes were described as a superposition of Hermite-Gaussians modes; as indicated in the S.I. This experimental overlap is performed in a similar way as \(\mathcal{F}\) from Equation~\ref{eq:FOM_em}, but not taking into account the phase information: exp. overlap \(\sim |\text{P}_x \cdot \text{E}_{m,x}| +|\text{P}_y \cdot \text{E}_{m,y}|\). The overlaps for the AP, RP and LP beams shown in Figure~\ref{fig:detailedStokes} and ~\ref{fig:RP_LP_LL_Q} are 0.92, 0.96 and 0.85, respectively.

To compare the lasing behavior of the different designs, we measured the light-in Vs light-out (L--L) curve for several nanolasers of each type, all of which have an exp. overlap \(\geq0.8\) and lasing wavelength between 655 nm and 695 nm (see S.I. for details). 
To take into account the variation of the lasing mode wavelength and its overlap with the gain spectrum of the QW, we compare the average L--L curve of the ensemble for each type of cavity in Figure~\ref{fig:RP_LP_LL_Q}(c). From fits to these curves, the lasing thresholds were found to be \(1.70 \pm 0.10\) nW (AP), \(1.63 \pm 0.14\) nW (RP), \(1.46 \pm 0.11\) nW (LP), which correspond to pulse energies of 17 pJ (AP), 16.3 pJ (RP), 14.6 pJ (LP), 4.66 pJ (Conv). These values are x3.1 -- x3.6 larger than the threshold shown by a conventional cavity of similar size: 0.466 \(\pm\) 0.032 nW.

The increase in input-output power conversion efficiency of inverse-designed nanolasers when compared to that of the conventional nanolasers was \(\times\)5.5 (AP), \(\times\)5.7 (RP), \(\times\)3.4 (LP) and the increase in maximum output power (at 8 nW input power) was \(\times\)4.9 (AP), \(\times\)5.1 (RP) and \(\times\)2.9 (LP) . This enhancement in the measured output power can be attributed to the enhancement in collection due to the axial emission obtained for the inverse-designed, as shown in Figure~\ref{fig:FoM_Epolar}(b).

The Q-factor of each cavity was obtained from from the photoluminescence spectrum (Figure~\ref{fig:RP_LP_LL_Q}(d)) by obtaining the Full Width Half Maximum, FWHM, of a gaussian fit to the lasing peak. The maximum value of the Q-factors were obtained for input powers near the lasing threshold, with values: 2365 (AP), 2181 (RP), 3250 (LP) and 2239 (Conv).

\section{Conclusions}
We have shown, using an adjoint-based topology optimization algorithm, that the output beam emitted by a whispering-gallery nanolaser can be tailored in terms of polarization and shape through the design of the contour of its cavity. The generality of this method has been demonstrated through the design of three cavities with different output beam shape and polarization, with the inverse-designed nanolasers exhibiting similar Q-factor and lasing threshold to conventional WGM lasers of comparable size. This control has been achieved within a footprint of less than 1 \(\mu\)m\(^2\) and opens up the possibility of producing monolithically integrated sub-micron laser sources with on-demand beam characteristics, which has potential in applications such as on-chip label-free biosensing \cite{Kim2018}, optical manipulation by integrated optical tweezers \cite{Cicek2020} and free-space optical communications \cite{Milione2015}.

\section{Methods}

\subsection{Fabrication procedure}
\label{section:fabrication}
The three cavity designs were fabricated on a III-V semiconductor platform with a 180-nm-thick GaInP/AlGaInP double quantum well on a GaAs substrate (EPSRC National Centre for III-V Technologies, Sheffield).

The full layer composition was:
10nm- Ga$_{0.51}$In$_{0.49}$P / 58nm- Al$_{0.357}$Ga$_{0.153}$In$_{0.49}$P / 10nm- Al$_{0.255}$GaIn$_{0.49}$P / 7nm- Ga$_{0.41}$In$_{0.59}$P / 10nm- Al$_{0.255}$GaIn$_{0.49}$P / 7nm- Ga$_{0.41}$In$_{0.59}$P / 10nm- Al$_{0.255}$GaIn$_{0.49}$P / 58nm- Al$_{0.357}$Ga$_{0.153}$In$_{0.49}$P / 10nm- Ga$_{0.51}$In$_{0.49}$P.

The semiconductor wafer was cleaved into 1 cm$^2$ chips. Each of the chips were cleaned with sequential ultrasonicated baths in acetone and isopropanol for 10 min each, and then blow-dried with nitrogen. The chips were spin-coated with TI-Prime and then with a 300 nm-thick layer of the negative-tone resist ma-N 2403, and crosslinked on a hot plate at 90 $^{\circ}$C for 2 min. The pattern was written by electron-beam lithography at 80 kV and beam current 1 nA with a Nanobeam nb4 system. Then the pattern was developed with MF-319 for 20 s and rinsed with deionized-water. The written pattern was transferred from the resist layer to the QW layer by Inductively-Coupled - Reactive Ion Etching (ICP-RIE) with the following conditions: gas-mixture 36 sccm Ar + 4 sccm Cl$_2$, chamber pressure 10 mTorr, RIE power 80 W, ICP power 700 W, sample stage at room temperature and etching time 1 min. The remaining resist was removed by an ultrasonicated bath in hot acetone. Finally, the pedestal under the cavities was formed by a selective wet-underetch with a hydrofluoric acid  solution (2.5\% w/w in deionized-water) for 1 min.

\subsection{Optical characterization}
\label{section:opticalCharacterization}
The nanolasers emission was characterized at room temperature via Fourier microscopy and k-space polarimetry. These two techniques allow the imaging of the nanolasers far-field intensity and polarization, respectively. A diagram of the optical setup is shown in Figure~\ref{fig:opticalSetup}.

The nanolasers were optically-pumped by a nanosecond-pulsed laser diode at 520 nm (NPL52B, Thorlabs) at a repetition rate of 100 Hz, with a pulse duration of 5 ns. This excitation was coupled through a dichroic beamsplitter (550 nm) into the objective lens (Nikon, x50, numerical aperture 0.5), which focused the excitation light into a $\sim$ 1 $\mu$m$^2$ diameter spot on the sample. The nanolaser emission was collected by the same objective. A  SPF and a LPF with cut-on/-off wavelength 550 nm were placed before and after the DM, respectively, to further spectrally filter the excitation and collection light.
The back-focal plane of the objective is directly imaged by a Fourier lens (FL, focal length = 20 cm), which is placed 20 cm away from the back focal plane of the objective, and projected by the infinitely-conjugated tube lens (TL, focal length = 5 cm) onto the fully-open entry slit of the spectrometer (Oxford Instruments - Kymera). The final image is projected onto a CCD camera (Andor iDus 416) by the spectrometer's diffraction grating at zero-order.

For the polarimetry analysis a quarter-wave plate (\(\lambda/4\)) and a linear polarizer (LP) were added in the beam path. Different orientations of these two optical components allowed the measurement of the Stokes parameters of the nanolasers emitted far-field. In addition, we needed to introduce a half-wave plate (\(\lambda/2\)) to rotate the linearly polarized light emerging after the LP to vertical polarization (perpendicular to the optical table) because the spectrometer grating has a larger scattering efficiency for this polarization.

For obtaining the photoluminescence spectrum of a nanolaser the FL is flipped out of the beam path and the spectrometer's grating (1200 l/mm, blaze 750 nm) is oriented to measure wavelengths within the photoluminescence wavelength range of the QW; from 650 nm to 700 nm. A neutral density (ND) filter was added after the source for varying the excitation power and acquire the L--L curves.

\begin{acknowledgement}
The authors thank S. Wedge for the use of the ICP-RIE at the David Bullet nanofabrication facilities at University of Bath and A. Elliot from the Geography department at University of Exeter for the access to the hydrofluoric acid etching facilities, S. Russo and K.J. Riisnaes for help with preliminary optical characterization experiments. This work was supported by the Engineering and Physical Sciences Research Council (Grant numbers EP/L015331/1 and EP/S001557/1).
\end{acknowledgement}

\begin{suppinfo}
The Supporting Information includes further details about: spatial filtering applied in the topology optimization method, Figure of Merit in the far-field, Stokes parameters, polarimetry analysis and experimental overlap calculation, characterization of all measured nanolasers, full layer structure of the quantum well Wafer.
\end{suppinfo}

\providecommand{\latin}[1]{#1}
\makeatletter
\providecommand{\doi}
  {\begingroup\let\do\@makeother\dospecials
  \catcode`\{=1 \catcode`\}=2 \doi@aux}
\providecommand{\doi@aux}[1]{\endgroup\texttt{#1}}
\makeatother
\providecommand*\mcitethebibliography{\thebibliography}
\csname @ifundefined\endcsname{endmcitethebibliography}
  {\let\endmcitethebibliography\endthebibliography}{}

\end{document}